\documentclass[12pt]{article}
\begin{document}

\begin{center}
{\Large  Classification of integrable discrete Klein-Gordon  models}

\vskip 0.2cm

{Ismagil T. Habibullin}\footnote{e-mail: habibullinismagil@gmail.com}\\

{Ufa Institute of Mathematics, Russian Academy of Science,\\
Chernyshevskii Str., 112, Ufa, 450077, Russia}\\

\bigskip

{Elena V. Gudkova}\footnote{e-mail: elena.gudkova79@mail.ru}

{Department of Applied Mathematics and Mechanics,\\ Ufa State Petroleum Technical University,
 \\Kosmonavtov str., 1, Ufa, 450062, Russia }\\

\bigskip

%\textbf{Abstract}

\end{center}
\begin{abstract}

The Lie algebraic integrability test is applied to the problem of classification of integrable Klein-Gordon type equations on quad-graphs. The list of equations passing the test is presented containing several well-known integrable models. A new integrable example is found, its higher symmetry is presented.
\end{abstract}
\bigskip


\section{Introduction}

We study the integrability problem for the quad--graph equation of the form
\begin{equation}\label{ddhyp}
u_{1,1}=f(u,u_1,  \bar u_{1} ).
\end{equation}
Such kind equations have a large variety of applications in physics, biology, architecture, etc. Here the unknown $u=u(m,n)$ is a function of two discrete variables $m,n$. For the sake of convenience we use the following notations: $u_k=u(m+k,n)$, $\bar{u}_k=u(m,n+k)$, $u_{1,1}=u(m+1,n+1)$.
Function $f$ is supposed to be locally analytic, it depends essentially on all three arguments. In other words equation (\ref{ddhyp}) can be rewritten in any of the following forms
\begin{equation}\label{ij}
u_{i,j}=f^{i,j}(u,u_i, \bar u_{j}),\label{i,j}\quad \mbox{with}\quad i=\pm1,\,j=\pm1.
\end{equation}
Nowadays various approaches are known for studying integrable discrete phenomena. The property of consistency around a cube \cite{NijhoffWalker}, has been proposed as the integrability criterion for quadrilateral difference equations \cite{BobenkoSuris}, \cite{Nijhoff}, \cite{abs}. Symmetry approach to the classification of integrable systems is adopted to discrete case  \cite{YL}, \cite{Xenitidis}, \cite{RasinHydon}, \cite{Mikhailov}, \cite{tongas}. Another characteristic property of an integrable equation is the vanishing of its algebraic entropy \cite{BellonViallet}. Alternative methods are used in \cite{NijhoffRamani}, \cite{GKP}, \cite{Hietarinta}.
In this article we study discrete phenomena from some different point of view. 

Years ago it was observed that characteristic Lie algebras, introduced in \cite{Shabat}, in the case of integrable hyperbolic type PDE's like sine-Gordon and Tzitzeica-Zhiber-Shabat equations have a very specific property. The dimensions of the linear spaces spanned by multiple commutators of the generators grow essentially slower than in generic case. In \cite{ZhiberMur} the problem of rigorous formalization of the notion of "slow growth" has been discussed. A conjecture was suggested  and checked by applying to classify integrable equations of the form $u_{x,y}=f(u,u_x).$

In the recent article \cite{HG} we introduced and successfully tested a classification scheme (called algebraic test) based on investigation of multiple commutators of characteristic vector fields defined by equation (\ref{ddhyp}). Now we use the algebraic test (an explanation is given in Section 2 below) to classify discrete Klein-Gordon type equation on quad graph
\begin{equation}\label{dKleinGordon}
u_{1,1}+u=g(u_1+\bar{u}_1).
\end{equation}
The list of equations which partially passed the algebraic test is studied additionally by applying the symmetry test. It is remarkable that the final list contains in addition to the well-known equations as the discrete potential KdV equation, the discrete Liuoville equation and a discrete analogue of the sine-Gordon equation also a new integrable model (see below the discussion on Corollary of theorem 2 and theorem 3): 
\begin{equation}\label{example}
u_{1,1}=\frac{\alpha u_1\bar u_1-\beta}{u(\alpha+u_1\bar u_1)}.
\end{equation}
Remind that the problem of complete description of integrable Klein -Gordon equations
\begin{equation}\label{KleinGordon}
u_{x,y}=g(u)
\end{equation}
has been solved years ago by A.V.Zhiber and A.B.Shabat \cite{ZS}. The authors proved that the only integrable equations in (\ref{KleinGordon}) are 
\begin{enumerate}
\item[] linear equation;
\item[] the Liouville equation $u_{x,y}=e^{u}$; 
\item[] the sine-Gordon equation $u_{x,y}=\sin u$;
\item[] the Tzitzeica-Zhiber-Shabat equation $u_{x,y}=e^{u}+e^{-2u}$.
\end{enumerate}

The article is organized as follows. In section 2 we define characteristic vector-fields and the test Lie ring, formulate the algebraic test, consider an illustrative example. The classification problem for the model of the form $u_{1,1}+u=g(u_1+\bar u_{1})$ passing the test is investigated in section 3. The  result of classification is summarized in Theorems 3 and 5.

\section{Algebraic test and classification scheme}

Note that for any integers $p$ and $q$ the variable $u_{p,q}$ is expressed in terms of the variables $\left\{u_i,\bar u_j\right\}_{i,j=-\infty}^{\infty}$ in a recurrent way. Hence the variables $u_i,\bar u_j$ are called dynamical variables. They considered as independent ones. For example, 
$$ u_{2,1}=f(u_1,u_2,u_{1,1})=f(u_1,u_2,f(u,u_1,\bar u_1)),$$
and
$$ u_{-1,-2}=f^{-1,-1}(\bar u_{-1},u_{-1,-1},\bar u_{-2})=f^{-1,-1}(\bar u_{-1},f^{-1,-1}(u, u_{-1},\bar u_{-1}),\bar u_{-2}),$$
  
Introduce the shift operators $D$ and $\bar D$, shifting the first and respectively the second integer arguments: $Dh(m,n)=h(m+1,n),$ $\bar Dh(m,n)=h(m,n+1)$.
Explain the action of the operators $D$ and $\bar D$ on the functions of dynamical variables. For the function $h=h(u_i,u_{i-1}...,u_{i'},\bar u_j,\bar u_{j+1},...,\bar u_{j'})$ we have 
$$h_k=D^kh=h(u_{i+k},u_{i+k-1}...,u_{i'+k},\bar u_{k,j},\bar u_{k,j+1},...,\bar u_{k,j'})$$
and similarly
$$\bar h_k=\bar D^kh=h(u_{i,k},u_{i,k-1}...,u_{i',k},\bar u_{k+j},\bar u_{k+j+1},...,\bar u_{k+j'}).$$
In these two formulas one has to replace the double shifts $u_{\alpha,\beta}$ through dynamical variables as it is discussed above.

Characteristic vector fields for the hyperbolic type PDE were introduced by E.Goursat \cite{Goursat} in 1899. They provide a very effective tool for classification of Liouville type integrable systems. For instance in \cite{Goursat} a list (almost complete) of hyperbolic type PDE was found admitting integrals in both directions by using a method based on the notion of characteristic vector fields. Interest to the subject renewed after the paper \cite{Shabat}, where the exhaustive description of exponential type systems is given admitting complete set of integrals. The concept of characteristic vector fields was adopted to the quad graph equations in \cite{hab}.
In the recent papers \cite{HabibullinPekcan}, \cite{HZhP}, \cite{HZhP2} the characteristic vector fields were used for the purpose of classification of semi-discrete Liouville type equations. All of these studies concern with the Liouville type integrability.  In \cite{ZhiberMur} and \cite{HG} some new applications of these important notions are suggested. Let us remind some necessary definitions. At first suppose that equation (\ref{ddhyp}) admits a non-trivial $n$-integral, i.e. there is a function $I$ depending on a finite number of the dynamical variables $I=I(u_{-j},u_{-j+1},...u_k)$ satisfying the equation $\bar DI=I$. In an enlarged form the last equation is
\begin{equation}\label{I}
I(r_{-j+1},r_{-j+2},...,r,\bar u,f,f_1,...f_{k-1})=I(u_{-j},u_{-j+1},...u_k),
\end{equation}
where the function $r=f^{-1,1}(u,u_{-1},\bar u_1)$ is defined in (\ref{ij}). Since the right hand side of (\ref{I}) does not depend on the variable $\bar u_{1}$ we find $\frac{\partial}{\partial \bar u_1}\bar DI=\frac{\partial}{\partial \bar u_1}\bar I=0$. It implies the equation $YI=0$ where the operator $Y$ is defined as follows $Y:=\bar D^{-1}\frac{\partial}{\partial \bar u_1}\bar D $.  Direct computations give (see \cite{HG},  \cite{hab})
\begin{equation}\label{Y}
Y=\frac{\partial}{\partial u}+x\frac{\partial}{\partial  u_{1}}+\frac{1}{x_{-1}}\frac{\partial}{\partial  u_{-1}}+
xx_1\frac{\partial}{\partial  u_{2}}+\frac{1}{x_{-1}x_{-2}}\frac{\partial}{\partial u_{-2}}+ ...,
\end{equation}
where $x=\bar D^{-1}(\frac{\partial f(u,u_1,\bar u_1)}{\partial \bar u_{1}})=-\frac{\partial f^{1,-1}(u,u_1,\bar u_{-1})/\partial u}{\partial f^{1,-1}(u,u_1,\bar u_{-1})/\partial u_{1}}$. We call the operators $X:=\frac{\partial}{\partial \bar u_{-1}}$ and $Y$ characteristic vector fields. Now it is evident that the map $f(u,u_1,\bar u_1)\Longrightarrow Y$ is correctly defined for any $f$ due to the formula (\ref{Y}).

Denote through $T$ the set of vector fields obtained by taking all possible multiple commutators of the operators $X$ and $Y$,  and taking all linear combinations with coefficients depending on a finite number of the dynamical variables $\bar u_{-1}$, $u$, $u_{\pm 1}$, $u_{\pm 2}$, ... . Evidently the set $T$ has a structure of the Lie ring. We call it test ring of the equation (\ref{ddhyp}) in the direction of $n$. In a similar way one can define the test ring $\bar T$ in the direction of $m$.

Notice that for Liouville type integrable equations of the form (\ref{ddhyp}) both rings $T$ and $\bar T$ are of finite dimension. Actually the test ring is a subset of the characteristic Lie ring \cite{hab}-\cite{HZhP2}.

Denote through $V_j$ the linear space over the field of locally analytic functions  spanned by $X$, $Y$ and all multiple commutators of $X$ and $Y$ of order less than or equal to j such that:
$$V_0=\{X,Y\},\quad V_1=\{X,Y,[X,Y]\}, \quad \dots .$$
Remind that such kind sequences of linear spaces have important applications in geometry\footnote{The authors thank E.Ferapontov for drawing their attention to this fact.} (see \cite{Cartan}, \cite{DoubrovZelenko}).
Introduce the function $\Delta(k)=\dim V_{k+1}-\dim V_k.$ The following conjecture is approved by numerous examples.

{\bf Conjecture (algebraic test)}. {\it Any integrable model of the form (\ref{ddhyp}) satisfies the following condition: there is a sequence of natural numbers $\left\{t_k\right\}_{k=1}^{\infty}$ such that $\Delta(t_k)\leq1$}.

Ring $T$ admits an automorphism, generated by the shift operator $D$, 
\begin{equation}\label{aut}
T\ni Z\stackrel{Aut}{\rightarrow} DZD^{-1}\in T,
\end{equation}
which plays crucial role in our further considerations. It is important that $X$ and $Y$ considered as operators on the set of functions depending on the variables $\bar u_{-1},u,u_{\pm1}, u_{\pm2},...$ satisfy the following conjugation relations \cite{HG}
\begin{equation}\label{autXY}
DXD^{-1}=pX \quad\mbox{and}\quad DYD^{-1}=\frac{1}{x}Y,
\end{equation} 
where $p=D(\frac{\partial f^{-1,-1}(u,u_{-1},\bar u_{-1})}{\partial \bar u_{-1}})=\frac{1}{\partial f^{1,-1}(u,u_1,\bar u_{-1})/\partial \bar u_{-1}}$. Indeed, specify the coefficients of the operator 
$DXD^{-1}=\sum a_i\frac{\partial}{\partial u_i}+p\frac{\partial}{\partial \bar u_{-1}}$
by applying it to the dynamical variables and find that $a_i=DXD^{-1}u_i=0$ for any integer $i$. Moreover $p=DXD^{-1}\bar u_{-1}=DXf^{-1,-1}(u,u_{-1},\bar u_{-1})=D(\frac{\partial f^{-1,-1}(u,u_{-1},\bar u_{-1})}{\partial \bar u_{-1}})$. In a similar way one can prove the second formula. Really apply the operator $DYD^{-1}=\sum c_i\frac{\partial}{\partial u_i}+d\frac{\partial}{\partial \bar u_{-1}}$ to $u_i$
and find $c_j=D(Yu_{j-1})=Yu_j.$ Then evaluate $d=DYD^{-1}\bar u_{-1}=f^{-1,-1}_u+\frac{1}{x_{-1}}f^{-1,-1}_{u_{-1}}$. Since $u_{-1,-1}=f^{-1,-1}(u,u_{-1},\bar u_{-1})$ one gets the equation $u=f^{-1,-1}(f(u,u_1,\bar u_1),u_1,\bar u_1)$. Let us differentiate  it with respect to $\bar u_{1}$ and find
$D\bar D(\frac{\partial f^{-1,-1}}{\partial u})\frac{\partial f}{\partial \bar u_{1}}+D\bar D(\frac{\partial f^{-1,-1}}{\partial u_{-1}})=0$ or the same $\frac{\partial f^{-1,-1}}{\partial u}+ \frac{1}{D^{-1}\bar D^{-1}(\frac{\partial f}{\partial \bar u_{1}})}\frac{\partial f^{-1,-1}}{\partial u_{-1}}=0$. Now due to the equation $x=\bar D^{-1}(\frac{\partial f}{\partial \bar u_{1}})$
 one concludes that $d=0$.

{\bf Lemma} 1. {\it Suppose that $Z=\sum_{-\infty}^{\infty} b_j\frac{\partial}{\partial  u_{j}}\in T$ satisfies the following two conditions: 
\begin{enumerate}
\item[1)] $DZD^{-1}=cZ$ for some function $c$;
\item[2)] $b_{j_0}\equiv 0$ for some fixed value of $j=j_0$. 
\end{enumerate}
Then $Z=0.$}

Proof of the Lemma can be found in \cite{HG}. 

{\bf Example}. As an illustrative example we consider the following discrete Liouville type equation (see \cite{AdlerStartsev})
\begin{equation}\label{as}
u_{1,1}=\frac{1}{u}(u_1-1)(\bar u_1-1).
\end{equation}
Find first an explicit form of the characteristic vector field $Y$. Since $f=\frac{1}{u}(u_1-1)(\bar u_1-1)$ and $\frac{\partial f}{\partial \bar u_{1}}=\frac{1}{u}(u_1-1)$ then 
\begin{equation}\label{as2}
x=\bar D^{-1}\frac{1}{u}(u_1-1)=\frac{1}{\bar u_{-1}}(u_{1,-1}-1).
\end{equation}
Express $u_{1,-1}$ through the dynamical variables due to equation (\ref{as}). Apply the operator $\bar D^{-1}$ to both sides of (\ref{as}) and find $u_1=\frac{1}{\bar u_{-1}}(u_{1,-1}-1)(u-1)$. Comparison of the last equation with (\ref{as2}) yields $x=\frac{u_1}{u-1}$, therefore 
\begin{equation}\label{Yas}
Y=\frac{\partial}{\partial u}+\frac{u_1}{u-1}\frac{\partial}{\partial  u_{1}}+\frac{u_{-1}-1}{u}\frac{\partial}{\partial  u_{-1}}+
\frac{u_1u_2}{(u-1)(u_1-1}\frac{\partial}{\partial  u_{2}}+ ...,
\end{equation}
For this equation $p=\frac{u-1}{u_1}=\frac{1}{x},$ hence $DXD^{-1}=pX$, $DYD^{-1}=pY$. Evaluate $D[X,Y]D^{-1}=p^2[X,Y]+pX(p)Y-pY(p)X$, since $X(p)=Y(p)=0$ we get $D[X,Y]D^{-1}=p^2[X,Y]$, moreover $[X,Y]=\sum_{j=1}^{\infty}a_j\frac{\partial}{\partial  u_{j}}+ a_{-j}\frac{\partial}{\partial  u_{-j}}$. Now due to the Lemma 1 one obtains $[X,Y]=0.$
So dimension of the ring $T$ for equation (\ref{as}) equals two and $\Delta(k)=0$ for all $k\geq0$ therefore equation (\ref{as}) passes the test.

\section{Equations of the form $u_{1,1}+u=g({u_1+\bar u_1})$}

In this section we apply the {\bf Conjecture} to the following Klein-Gordon type particular class of discrete model (\ref{ddhyp})
\begin{equation}\label{pcase}
u_{1,1}+u=g({u_1+\bar u_1}).
\end{equation}
where the function $g$ is to be determined. 

{\bf Classification scheme}. Obviously equation (\ref{ddhyp}) passing the algebraic test above should satisfy one of the conditions:
\begin{enumerate}
\item[i)] $\Delta(0)<\Delta_{max}(0)=1$;
\item[ii)] $\Delta(0)=\Delta_{max}(0)$, $\Delta(1)<\Delta_{max}(1)=2$; 
\item[iii)] $\Delta(0)=\Delta_{max}(0)$, $\Delta(1)=\Delta_{max}(1)$, $\Delta(2)<\Delta_{max}(2)=3$;
\item[iv)] $\Delta(0)=\Delta_{max}(0)$, $\Delta(1)=\Delta_{max}(1)$, $\Delta(2)=\Delta_{max}(2)$ and  $\Delta(k)\leq1$ for some $k>2$.
\end{enumerate}
Where $\Delta_{max}(k)$ stands for the greatest value of $\Delta(k)$ for equation (\ref{ddhyp}) when $f(u,u_1,\bar u_{1})$ ranges the class of arbitrary functions.

Remark that in the case of equation (\ref{pcase}) investigation of the first three particular cases $i)$-$iii)$ allows one to extract a very short list of equations expected to be integrable. The list is exhaustive because the case $iv)$ is never realized (see below Corollaries of Theorems 2 and 4).

Introduce vector fields $R_1=[X,Y]$, $P_1=[X,R_1],$ $Q_1=[Y,R_1]$, $R_2=[X,Q_1]$, $W=[Y,Q_1]$, $Z=[X,P_1]$. Using these vector fields we can span in addition to $V_0$ and $V_1$ two more linear spaces:
$$V_2=V_1+\{P_1,Q_1\},\quad V_3=V_2+\{W,Z,R_2\}.$$

In order to evaluate $\Delta (k)$ we will use the automorphism (\ref{aut}). At first evaluate the factors $x$ and $p$ in formula (\ref{autXY}) for the case (\ref{pcase}). We have $x=p=g'(g^{-1}(u_1+\bar u_{-1}))$, where the function  $\beta=g^{-1}(\alpha)$ is the inverse to the function $\alpha=g(\beta)$. Inversely, knowing $x=x(u_1+\bar u_{-1})$ one can recover $g(\beta)$ by using the equation
\begin{equation}\label{restoreg}
\beta=g^{-1}(\alpha)=\int (g^{-1}(\alpha))'d\alpha=\int\frac{d\alpha}{g'(g^{-1}(\alpha)}=\int\frac{d\alpha}{x(\alpha)}.
\end{equation}
Specify the action of the characteristic operators $X$ and $Y$ on the variable $x$. Evidently, $Xx=x'$, $Yx=xx'$. It is found by direct calculation that
\begin{eqnarray}
DR_1D^{-1}&=&R_1-\frac{x'}{x}Y-x'X,\nonumber\\
DP_1D^{-1}&=&xP_1-x'R_1-rY-xrX,\quad r=x^{''}-\frac{x^{'2}}{x},\nonumber\\
DQ_1D^{-1}&=&\frac{1}{x}Q_1+\frac{x'}{x}R_1-\frac{x^{''}}{x}Y-x^{''}X,\nonumber\\
DWD^{-1}&=&\frac{1}{x^2}W+(\frac{2x^{''}}{x}-\frac{x^{'2}}{x^2})R_1-\frac{x^{'''}}{x}Y-x^{'''}X,\label{tableconjec}\\
DZD^{-1}&=&x^2Z+(x^{'2}-2xx^{''})R_1-qY-xqX,\quad q=xx^{'''}-2x^{'}x^{''}+\frac{x^{'3}}{x},\nonumber\\
DR_2D^{-1}&=&R_2-\frac{x'}{x}Q_1+x'P_1-\frac{x^{'2}}{x}R_1-sY-xsX,\quad s=x^{'''}-\frac{x'x^{''}}{x},\nonumber
\end{eqnarray}

Let us study the set $G$ of all multiple commutators of $X$ and $Y$.

{\bf Lemma 2}. {\it The coefficients of any operator in $G$ are functions of a finite number of the dynamical variables $x$, $x_{\pm1}$, $x_{\pm2},...\,.$}

{\bf Proof}. Here $x(\alpha)$ is a fixed function of the argument $\alpha=u_1+\bar{u}_{-1}$. Therefore one can write $x'=\phi(x)$ for some function $\phi$. Then $X(x)=-\phi(x)$ and 
$Y(x)=x\phi(x)=:\psi(x)$. By using the conjugation relations $DXD^{-1}=xX$ and $DYD^{-1}=\frac{1}{x}Y$ one derives that $X(x_j)=\phi^j(x,x_1,...x_j)$ and $Y(x_j)=\psi^j(x,x_1,...x_j)$. Similarly $X(x_{-j})=\phi^{-j}(x,x_{-1},...x_{-j})$ and $Y(x_{-j})=\psi^{-j}(x,x_{-1},...x_{-j})$. Now evidently $R_1=X(x)\frac{\partial}{\partial u_1}+X(\frac{1}{x_{-1}})\frac{\partial}{\partial u_{-1}}+X(xx_1)\frac{\partial}{\partial u_2}+\cdots$ satisfies the statement of the lemma. Due to the formulas $R_1(x)=X(x)x'$ and $DR_1D^{-1}=R_1+\frac{x'}{x}Y-x'X$ one gets $R_1(x_j)=\phi^j(x,x_1,...x_j)$. Obviously the proof can be completed by using induction.

{\bf Theorem 2}. {\it Suppose that an equation of the form (\ref{pcase}) satisfies one of the conditions $i)-iii)$ of the {\bf Classification scheme} then function $x=x(\alpha)$ solves the following ordinary differential equation
\begin{equation}\label{ode}
x^{'2}=(x^2+1)\gamma+x\nu
\end{equation}
with the constant coefficients $\gamma, \nu$.}

{\bf Proof of the theorem 2}. Begin with the case $i)$, suppose that $\Delta (0)=0$, then we have $R_1=\lambda X +\mu Y$. It is evident that $R_1=X(x)\frac{\partial}{\partial  u_{1}}+...$, $X=\frac{\partial}{\partial \bar u_{-1}}$ and $Y=\frac{\partial}{\partial  u}+x\frac{\partial}{\partial  u_{1}}+...$ hence $\lambda=\mu=0$ and therefore $R_1=0$.  By applying the automorphism above to both sides of the last equation one finds 
$$-\frac{x'}{x}Y-x'X=0.$$
Since $X$ and $Y$ are linearly independent we get equation $x'=0$ which is a particular case of (\ref{ode}). Evidently its solution is $x=c$ and due to (\ref{restoreg}) it can be found that $\beta=g^{-1}(\alpha)=\frac{1}{c}\alpha+c_1$. Thus our equation $\alpha=g(\beta)$ (see (\ref{pcase})) is linear $u_{1,1}+u=c(u_1+\bar u_1+c_1)$. For this case $\dim T=2$ so that $\Delta (k)=0$ for $k\geq0$. In a similar way one checks that condition $ii)$ leads to (\ref{ode}). Indeed  suppose that $\Delta(0)=1$ and $\Delta(1)<2$ then we have 
\begin{equation}\label{Delta1}
P_1=\nu Q_1+\epsilon R_1.
\end{equation}
Here due to Lemma 2 functions $\nu$ and $\epsilon$ might depend only on $x$, $x_{\pm 1}$, $x_{\pm 2}\dots$.
Apply the automorphism (\ref{aut}) to both sides of (\ref{Delta1}) then simplify due to equations (\ref{tableconjec}):
$$x(\nu Q_1+\epsilon R_1)- x'R_1-rY-xrX=D(\nu)(\frac{1}{x}Q_1+\frac{x'}{x}R_1-\frac{x^{''}}{x}Y-x^{''}X)+ D(\epsilon)(R_1-\frac{x'}{x}Y-x'X).$$
Comparison of the coefficients before linearly independent operators gives rise to the conditions 
\begin{eqnarray}
Q_1:&&\quad x\nu=\frac{1}{x}D(\nu); \nonumber\\
R_1:&&\quad -x'+ x\epsilon=\frac{x'}{x}D(\nu)+D(\epsilon); \nonumber\\
Y:&&\quad -r=-\frac{x'}{x}D(\epsilon)+\frac{x''}{x}D(\nu); \nonumber\\
X:&&\quad -xr=-x'D(\epsilon)-x''D(\nu). \nonumber
\end{eqnarray}
Simple analysis of these equations implies that equation (\ref{Delta1}) holds if and only if the following three conditions 
valid: $\nu=0$, $\epsilon=const,$ $x'=\epsilon(x-1).$ Actually, under these conditions the last two equations above are satisfied automatically. In a similar way one can check that $Q_1=\nu P_1-\epsilon R_1$ is equivalent to the same three conditions. Hence if $\Delta (1)<2$ then $\Delta (1)=0$ and consequently  $\Delta (k)=0$ for any natural $k\geq1$.  Thus in this case $\dim T=3$.

Let us suppose now that $\Delta (0)=1$, $\Delta (1)=2$ and $\Delta (2)\leq2$ which corresponds to the case $iii)$. At first consider the case when $Z$ is linearly expressed through the other vector fields in subspace $V_3$:
\begin{equation}\label{z}
Z=\gamma R_1+ \delta P_1+\epsilon Q_1+\phi R_2+\psi W.
\end{equation}
Apply the automorphism (\ref{aut}) to both sides of equation (\ref{z}) and compare coefficients before linearly independent operators
\begin{eqnarray}
W:&\,x^2\psi&=D(\psi)\frac{1}{x^2},\nonumber \\
R_2:&\,x^2\phi&=D(\phi),\nonumber \\
Q_1:&\,x^2\epsilon &=D(\epsilon)\frac{1}{x}+\frac{x'}{x}\phi,\nonumber \\
P_1:&\,x^2\delta &=D(\delta)x+x'\phi,\nonumber \\
R_1:&\, x^2\gamma &+x^{'2}-2xx^{''}=D(\delta)x'+D(\gamma).\nonumber 
\end{eqnarray}
Since $x=x(u_1+\bar u_{-1})$ we have $\psi=0$, $\phi=0$, $\epsilon=0$, $\delta=0$, $\gamma=const$. Comparison of the coefficients of $X$ and $Y$ gives one more equation $xq=\gamma x'$. Finally we get two ordinary differential equations for $x$ which are absolutely the same as in the case of the equation $u_{1,1}-u=g({u_1-\bar u_1})$ (see \cite{HG}):
$$x^2x^{'''}-2xx'x^{''}+x^{'3}=\gamma x', \quad (x^2-1)\gamma+x^{'2}-2xx^{''}=0. $$
The compatibility condition of these equations is equivalent to equation (\ref{ode}). In this case we have $Z=\gamma R_1$. It is remarkable that  $x$ solves equation (\ref{ode}) if and only if $W$ is linearly expressed through the other elements of $V_3$ and then $W=\gamma R_1$. And the last possibility is when $R_2$ is linearly expressed through $X$, $Y$, $R_1$, $P_1$, $Q_1$, $W$, $Z$. In this case $x$ solves the equation $x'=0$. The proof of the theorem is completed. 

In order to find $x=x(\alpha)$ evaluate the integral
\begin{equation}\label{x}
H(x):=\int\frac{dx}{\sqrt{(x^2+1)\gamma+x\nu}}=\alpha-\alpha_0. 
\end{equation}
For the case $\gamma\neq0$ the answer is given by the formula
\begin{equation}\label{H}
H(x)=\frac{1}{\sqrt{\gamma}}\ln(2\sqrt{x^2+1+xb}+2x+b), \quad b=\frac{\nu}{\gamma}.
\end{equation}
Now find $x$ by solving the equation $H(x)=\alpha-\alpha_0$. 
$$x(\alpha)=\frac{1}{4}e^{\sqrt{\gamma}(\alpha-\alpha_0)}-\frac{\nu}{2\gamma}-(1-\frac{\nu^2}{4\gamma^2})e^{-\sqrt{\gamma}(\alpha-\alpha_0)}.$$
In order to get the corresponding quad-graph equation (\ref{pcase}) integrate again:
\begin{equation}\label{g}\beta=g^{-1}(\alpha)=\int\frac{d\alpha}{x(\alpha)}.\end{equation}
Integration gives
\begin{equation}\label{gg}\beta=\frac{1}{\sqrt{\gamma}}\ln \left|\frac{e^{\sqrt{\gamma}(\alpha-\alpha_0)}-b-2}{e^{\sqrt{\gamma}(\alpha-\alpha_0)}-b+2}\right|+\beta_0.\end{equation}
Then the equation searched is given by the formula (take $\sqrt{\gamma}=1$)
$$a_1e^{u_{1,1}+u_1+\bar{u}_1+u}+a_2e^{u_{1,1}+u}+a_3e^{u_1+\bar{u}_1}+a_4=0,$$
where $a_1=e^{-\alpha_0-\beta_0}$, $a_2=-e^{-\alpha_0}$, $a_3=(2-b)e^{-\beta_0}$, $a_4=2+b$. By setting $u=\ln v$ one reduces it to a quadrilinear form
\begin{equation}\label{new}
a_1v_{1,1}v_1\bar{v}_1v+a_2v_{1,1}v+a_3v_1\bar{v}_1+a_4=0,
\end{equation}
with arbitrary constants $a_j$. 

If $\gamma=0$, then $x=x(\alpha)$ solves the equation $x'=\sqrt{\nu x}$. Thus obviously $x=c(\alpha-\alpha_0)^2$ and therefore 
$$\beta=g^{-1}(\alpha)=\int\frac{d\alpha}{c(\alpha-\alpha_0)^2}=\frac{d}{\alpha-\alpha_0}-\beta_0,\, d=-1/c.$$
The corresponding equation (\ref{pcase}) follows from $(\alpha-\alpha_0)(\beta-\beta_0)=d.$ It is of the form
\begin{equation}\label{dpkdv}
(u_{1,1}+u-\alpha_0)(u_1+\bar{u}_1-\beta_0)=d.
\end{equation}
 
{\bf Corollary of Theorem 2}. {\it If for a nonlinear chain (\ref{pcase}) one of the conditions $i)-iii)$ is satisfied then the chain is of one of the form:
\begin{enumerate}
\item[1)] $u_{1,1}+u=c(u_1+\bar u_1+c_1)$;
\item[2)] $a_1u_{1,1}u_1\bar{u}_1u+a_2u_{1,1}u+a_3u_1\bar{u}_1+a_4=0$; 
\item[3)] $(u_{1,1}+u-\alpha_0)(u_1+\bar{u}_1-\beta_0)=d$.
\end{enumerate}}

Let us study in details the list obtained. It is remarkable that it contains a new integrable example of equation (\ref{ddhyp}). 

{\bf Theorem 3}. {\it The chain 2) in the Corollary of Theorem 2 admits a symmetry of the form 
\begin{equation}\label{sym}
u_t=g(u_{1},u_{-1},u,\bar u_1,\bar u_{-1}) 
\end{equation}
if and only if the condition holds $k:=\frac{a_3}{a_2}=\pm1.$ Under this condition the symmetry is of the form
$$u_t=\lambda u\frac{ku_1+u_{-1}}{ku_1-u_{-1}}+\mu \bar u\frac{k\bar u_1+\bar u_{-1}}{k\bar u_1-\bar u_{-1}}.$$
Chain 3) admits a symmetry of the form (\ref{sym}) if and only if $\alpha_0=\beta_0$ and then it is reduced to the discrete potential KdV equation.}

Theorem can be proved by using technique developed in \cite{YL}. Actually it claims that the chain 2) with $a_3=\pm a_2$ passes the symmetry test. For the case $k=1$ (or $a_3=a_2$) the chain is a particular case of the Viallet equation (see \cite{BellonViallet}), this means that by a point transformation it can be reduced to one of the equations of ABS list \cite{abs}. Higher symmetries for this case are found  in \cite{Xenitidis}. For some special choice of the parameters chain 2) with $k=-1$ is reduced to the discrete sin-Gordon equation 
\begin{equation}\label{heredero}
3(u_{1,1}u_1\bar{u}_1u-1)+u_{1,1}u-u_1\bar{u}_1=0
\end{equation}
found by R. Hernandez Heredero and C. Scimiterna. Recently in \cite{YL} it was proved that equation (\ref{heredero}) admits higher symmetries. To the best of our knowledge general case of the equation 2) with $k=-1$ has never been identified before as integrable one. 

When $a_1=0$
and $a_3=-a_2$ it is reduced to the well-known discrete Liouville equation
$$e^{u_{1,1}+u}=e^{u_{1}+\bar{u}_1}+1,$$
having nontrivial integrals, i.e. functions solving the equations $\bar DF=F$ and $D\bar F=\bar F$, with $F=e^{u-u_1}+e^{u_2-u_1}$ and $\bar F=e^{u-\bar{u}_1}+e^{\bar{u}_2-\bar{u}_1}$.

The second part of the theorem claims that case 3) admits a symmetry of small order only if it coincides up to a point transformation with the discrete pkdv equation. The question is still open whether it admits any symmetry of a more complicated form, say
$u_t=g(u_{2},u_{1},u_{-1},u,\bar u_1,\bar u_{-1},\bar u_{-2})$. There are some technical difficulties with applications of the symmetry approach to discrete models (see discussion in \cite{YL}).

Turn back to the set $G$ consisting of $X$, $Y$ and their all multiple commutators. Assign two integers: order and degree to each element in $G$.  Define the order of an element $Z\in G$ as a number of its factors $X$ and $Y$ minus one. For instance $ord [X,Y]=1$,  $ord [X,[X,Y]]=2$ and so on. Define the degree $deg (Z)$ of $Z$ as the exponent $k$ in the expansion of the operator obtained by applying the automorphism (\ref{aut}): $DZD^{-1}=x^kZ+...$, where the tail is a linear combination of the elements with order less than $ord(Z)$. Denote through $G_{i,j}$ a subset of $G$ containing elements with the order $i$ and the degree $j$. Let $G_i=\bigcup_{j}G_{i,j}$ be the union of all $G_{i,j}$ with one and the same $i$. Evidently the set $G_{i,i-1}$ (as well as $G_{i,-i+1}$) contains the only element $Z_{i,i-1}=ad_X^i(Y)$ up to the factor $-1$ (correspondingly, the only element $Z_{i,-i+1}=ad_Y^i(X)$ up to the factor $-1$). Here the operator $ad$ is defined as follows $ad_X(Y)=[X,Y]$. 

{\bf Theorem 4}. {\it Suppose that $Z_{k,k-1}$ (or $Z_{k,-k+1}$) is in the basis of the linear space $V_k\supset G_k$ for any integer $k$: $3\leq k<N$, but $Z_{N,N-1}$ (respectively $Z_{N,-N+1}$) is linearly expressed through the other operators in $V_N$ then function $x=x(u_1+\bar u_{-1})$ solves an equation of the form $x'=\epsilon(x-1)$ with the constant coefficient $\epsilon$.}

The proof of the theorem is done exactly in the same way as the proof of Theorem 3 from \cite{HG}, but presented in the article for the readers' convenience.

{\bf Lemma} 3. {\it For any integer $k\geq 3$ we have
\begin{equation}\label{zconjugation}
DZ_{k+1,k}D^{-1}=x^kZ_{k+1,k}+c_kx'x^{k-1}Z_{k,k-1}+\cdots\, ,
\end{equation}
where $c_k\geq0$ (but $c_k>0$ for $k>3)$ and the tail contains a linear combination of the operators with the order less than $k$. }

Prove the Lemma 3 by induction.  From the list of equations 
(\ref{tableconjec}) one gets for $Z=Z_{3,2}$ and $R_1=Z_{1,0}$ the following representation
\begin{equation}\label{z32}
DZ_{3,2}D^{-1}=x^2Z_{3,2}+(x^{'2}-2xx^{''})Z_{1,0}-qY-xqX,\quad q=xx^{'''}-2x^{'}x^{''}+\frac{x^{'3}}{x}
\end{equation}
showing that the statement is true for the case $k=3$. Suppose now that $DZ_{k,k-1}D^{-1}=x^{k-1}Z_{k,k-1}+c_{k-1}x'x^{k-2}Z_{k-1,k-2}+\cdots\,$ and evaluate $DZ_{k+1,k}D^{-1}$:
$$DZ_{k+1,k}D^{-1}=[xX,x^{k-1}Z_{k,k-1}+c_{k-1}x'x^{k-2}Z_{k-1,k-2}+\cdots\,]=x^kZ_{k+1,k}+c_kx'x^{k-1}Z_{k,k-1}+\cdots\,$$
where $c_k=c_{k-1}+(k-1)>c_{k-1}\geq0$. The proof is completed.

{\bf Proof of the Theorem 4}. Suppose that 
\begin{equation}\label{decompositionzN}
Z_{N,N-1}=\sum_{ord (Z_\nu)=N}a_{\nu}Z_{\nu}+\sum_{ord (Z_\mu)=N-1}b_{\mu}Z_{\mu}+\cdots\, ,
\end{equation}
where $Z_\nu$ and $Z_\mu$ range the basis of $V_N$ and the tail contains a linear combination of the operators of less order.
Apply the automorphism (\ref{aut}) to both sides of (\ref{decompositionzN}): 
\begin{eqnarray}
&&x^{N-1}(\sum_{ord (Z_\nu)=N}a_{\nu}Z_{\nu}+\sum_{ord (Z_\mu)=N-1}b_{\mu}Z_{\mu}+\cdots)-x^{N-2}x'c_{N-1}Z_{N-1,N-2}+\cdots=\nonumber\\
&&=\sum_{ord (Z_\nu)=N}D(a_{\nu})(x^{k_{\nu}}Z_{\nu}+\cdots) +\sum_{ord (Z_\mu)=N-1}D(b_{\mu})(x^{k_{\mu}}Z_{\mu}+\cdots). \nonumber
\end{eqnarray}
Collect the coefficients before $Z_{\nu}$ and get 
\begin{equation}\label{comparison}
x^{N-1}a_{\nu}=x^{k_{\nu}}D(a_{\nu}), \quad k_{\nu}\neq N-1.
\end{equation}
Due to Lemma 2 functions $a_{\nu}$ and $b_{\mu}$ depend on $x$ and its shifts. Moreover it follows from (\ref{comparison}) that $a_{\nu}$ cannot depend on $x$, $x_{\pm 1}$, $x_{\pm 2},\dots$ at all. Therefore the only possibility is $a_{\nu}=0$. Compare now the coefficients before $Z_{N-1,N-2}$ and find:
\begin{equation}\label{comparison2}
x^{N-1}b-c_{N-1}x^{N-2}x'=x^{N-2}D(b), 
\end{equation}
where $b$ is the coefficient of $Z_{N-1,N-2}$ in the expansion (\ref{decompositionzN}). A simple analysis of equation (\ref{comparison2}) shows that $b$ is constant. Thus (\ref{comparison2}) is equivalent to the equation $x'=\epsilon(x-1)$ with $\epsilon=b/c_{N-1}$.

{\bf Corollary of Theorem 4}. {\it The case $iv)$ of the {\bf Classification scheme} is never realized.}

{\bf Proof}. Suppose on contrary that such a case is realized. Then at least one of the vector fields $Z_{k,k-1}$ or $Z_{k,-k+1}$ should be linearly expressed through other elements of $V_k$, otherwise $\Delta(k)\geq2$. Therefore due to Theorem 4 we have $x'=\epsilon(x-1)$ which corresponds to the case $ii)$ $\Delta(0)=1$, $\Delta(1)<2$ and $\dim T=3$. The contradiction shows that our assumption is not true. The proof is completed.

Let us summarize the results of reasonings  above in the following theorem.

{\bf Theorem 5}. {\it Suppose that an equation of the form (\ref{pcase}) passes the algebraic test then it is of one of the form given in Corollary of Theorem 2.}

\section{Conclusions.}

A classification scheme introduced recently in \cite{HG} is applied to quad-graph Klein-Gordon type equation. The list of equations passed the test contains along with the well-known integrable models some new example which passes also the symmetry test.

\section*{Acknowledgments}
The authors thank Prof. A.V.Zhiber and Dr. R.N.Garifullin for valuable advises. This work is partially supported by Russian Foundation for Basic
Research (RFBR) grants $\#$ 10-01-91222-CT-a, $\#$
11-01-00732-a, $\#$
11-01-97005-r-povoljie-a, and $\#$ 10-01-00088-a.

\end{document}